\pdfoutput=1
\documentclass[10pt, twocolumn]{article}
\usepackage[utf8]{inputenc}

\usepackage{times}
\usepackage{verbatim}
\usepackage{graphicx} 
\usepackage{algpseudocode}
\usepackage{booktabs}
\usepackage[font=bf]{caption}

\usepackage[utf8]{inputenc}
\usepackage{algorithm}

\usepackage[hyphens]{url}

\usepackage{comment}
\usepackage{authblk}
\usepackage{listings} 
\newcommand{\code}[1]{\texttt{#1}}

\usepackage[svgnames]{xcolor}
  \definecolor{diffstart}{named}{Grey}
  \definecolor{diffincl}{named}{Green}
  \definecolor{diffrem}{named}{OrangeRed}

\definecolor{codegreen}{rgb}{0,0.6,0}
\definecolor{codegray}{rgb}{0.5,0.5,0.5}
\definecolor{codepurple}{rgb}{0.58,0,0.82}
\definecolor{backcolour}{rgb}{0.95,0.95,0.92}

\lstdefinelanguage{Python}{
    basicstyle=\ttfamily\footnotesize,
    morecomment=[f][\color{diffstart}]{@@},
    morecomment=[f][\color{diffincl}]{+\ },
    morecomment=[f][\color{diffrem}]{-\ },
    morecomment=[f][\color{teal}]{>\ },
    morecomment=[f][\color{purple}]{>>\ },
    keywords={def,if,for,in, self, print},
}

\lstdefinestyle{mystyle}{
    language=Python,
    backgroundcolor=\color{backcolour},   
    commentstyle=\color{codegreen},
    keywordstyle=\color{magenta},
    numberstyle=\tiny\color{codegray},
    stringstyle=\color{codepurple},
    basicstyle=\ttfamily\footnotesize,
    breakatwhitespace=false,         
    breaklines=true,                 
    captionpos=b,                    
    keepspaces=true,                 
    numbers=left,                    
    numbersep=5pt,                  
    showspaces=false,                
    showstringspaces=false,
    showtabs=false,                  
    tabsize=2
}

\title{Q-PAC: Automated Detection of Quantum Bug-Fix Patterns}

\author[1]{Pranav K. Nayak}
\author[1]{Krishn V. Kher}
\author[1]{M. Bharat Chandra}
\author[1]{M. V. Panduranga Rao\vspace{-1ex}}
\author[2]{Lei Zhang}
\affil[1]{Department of Computer Science and Engineering, IIT Hyderabad, India}
\affil[2]{Department of Information Systems, University of Maryland, Baltimore County, USA}

\date{}
 
\begin{document}
\twocolumn[
    \begin{@twocolumnfalse}
        \maketitle
        \begin{abstract}
    \textbf{Context}: Bug-fix pattern detection has been investigated in the past in the context of classical software. However, while quantum software is developing rapidly, the literature still lacks automated methods and tools to identify, analyze, and detect bug-fix patterns. To the best of our knowledge, our work previously published in SEKE'23 was the first to leverage classical techniques to detect bug-fix patterns in quantum code.

\textbf{Objective}: To extend our previous effort, we present a research agenda (Q-Repair), including a series of testing and debugging methodologies, to improve the quality of quantum software. The ultimate goal is to utilize machine learning techniques to automatically predict fix patterns for existing quantum bugs.

\textbf{Method}: As part of the first stage of the agenda, we extend our initial study and propose a more comprehensive automated framework, called Q-PAC, for detecting bug-fix patterns in IBM Qiskit quantum code. In the framework, we develop seven bug-fix pattern detectors using abstract syntax trees, syntactic filters, and semantic checks.

\textbf{Results}: To demonstrate our method, we run Q-PAC on a variety of quantum bug-fix patterns using both real-world and handcrafted examples of bugs and fixes. The experimental results show that Q-PAC can effectively identify bug-fix patterns in IBM Qiskit. 

\textbf{Conclusion}: We hope our initial study on quantum bug-fix detection can bring awareness of quantum software engineering to both researchers and practitioners. Thus, we also publish Q-PAC as an open-source software on GitHub. We would like to encourage other researchers to work on research directions (such as Q-Repair) to improve the quality of the quantum programming.\\
        \end{abstract}
    \end{@twocolumnfalse}
]
\section{Introduction}
Quantum computing has the potential to revolutionize computing across multiple domains, including artificial intelligence, optimization, healthcare, energy, and space. Particularly exciting 
is the prospect of \emph{quantum advantage}---provable superiority over classical computation in terms of resources like time and space. Recent breakthroughs~\cite{daley2022practical,huang2022quantum} have brought applications of quantum computing from just a promising future to a present reality. Quantum software development is essential for realizing quantum advantage in applications by creating algorithms, tools, and techniques that leverage the unique properties of quantum systems.

With the increasing size and complexity of quantum programs being written, it is natural to expect an increased number of bugs and for more complicated bugs to creep into quantum source code. Indeed, this phenomenon is folklore in classical software~\cite{IncorrProg}. Therefore, a significant body of research, tools, and techniques exist in the detection, analysis and elimination of bugs in classical software~\cite{BugTypes}. These techniques range from static code analysis~\cite{StatAn} to run-time detection~\cite{testing}. \\
$~~~~$The effort to understand and classify commonly occurring bugs yields rich dividends. Steps for this approach include the identification and classification of bug patterns, the design of bug fixes, and the detection of bug-fix patterns. Correct identification of bug-fix patterns is of immense use in statistical analysis of bugs, their prevalence, and fixes. This helps in streamlining and developing tools for automatic bug detection, fixing, and manpower training. In this paper, we are particularly interested in automated approaches to detecting bug-fix patterns in quantum programs. 

In classical software engineering, common bug-fix patterns are well studied, and both manual and automated approaches have been proposed~\cite{campos2017common,Madeiral2018,pan2009toward,BugDataSet}. In quantum software engineering~\cite{Piattini}, preliminary studies exist in testing of quantum programs~\cite{miranskyy2019testing,miransky2020bug,miranskyy2021testing}, and bug patterns of quantum programs~\cite{huang2019statistical,zhao2020quantum,zhao2021identifying}. 

In this paper, we pose a research agenda for automated quantum software bug fixing, and extend our initial work on bug-fix pattern detection~\cite{Kher23} to develop a more sophisticated tool called Quantum PAttern Classifier (\emph{Q-PAC}). The tool uses abstract syntax trees (AST)~\cite{AST} and regular expressions (regex) to compare buggy and patched files by extracting information specific to bug-fix patterns (see Section 3). The detectors in the tool then positively identify or reject the bug-fix pattern. The tool currently has the ability to classify seven quantum bug-fix patterns in the context of Qiskit~\cite{Qiskit:online} code. In accordance with the research agenda, this tool will be further developed and will incorporate 
machine learning techniques for automated fix pattern prediction.

This extended version differs from the preliminary version~\cite{Kher23} in the following important ways:
\begin{enumerate}
\item We propose a research agenda called \emph{Q-Repair} to extend the existing efforts of bug and fix pattern detection techniques in quantum software engineering. The ultimate goal of \emph{Q-Repair} is to predict fix patterns automatically based on existing bugs.
\item The architecture and code of~\cite{Kher23} have been refactored, generalizing it and making it easily extensible.\footnote{Our source code is publicly available at \url{https://github.com/pranavknayak/Q-PAC}.} The tool is optimized for performance in that it generates the ASTs only once for a buggy-fixed code pair. 
A coarse filter that prunes the search space of patterns has been incorporated.
\item While multiple patterns apply to a single bug-fix pair, the preliminary version only reports one of the patterns; in contrast, \emph{Q-PAC} now reports all patterns detected.
\item In~\cite{Kher23}, we left the problem of detecting patterns spanning multiple lines as an open one; in this paper, we design detectors for some such patterns.
\item More bug-fix patterns have been added as compared to~\cite{Kher23} (four more, to be precise), to illustrate the wide applicability of the tool.
\end{enumerate}

This paper is arranged as follows. Section~\ref{sec:related} discusses some related work, including bug patterns and bug-fix patterns in both classical and quantum contexts. In Section~\ref{sec:method}, we introduce our method to deploy bug-fix pattern detection to automated fix pattern predictions. Section~\ref{sec:framework} outlines our proposed framework for bug-fix pattern detections. We present the detailed implementation of \emph{Q-PAC} in Section~\ref{sec:qdiff}. Section~\ref{sec:threats} discusses the threats to validity. We then conclude this paper with a discussion of future directions in Section~\ref{sec:conclusion}.

\section{Related Work}\label{sec:related}
In the interest of space, we assume a working knowledge of quantum computing~\cite{nielsen_chuang_2010}. Here, we briefly discuss
some relevant points.
Quantum computing is a paradigm of computation that seeks to harness phenomena peculiar to quantum mechanics, to speed up computation~\cite{nielsen_chuang_2010}.
 A very popular model of quantum computation is the so-called \emph{circuit} model, where computation proceeds through the
 application of a sequence of \emph{unitary gates} and \emph{measurements}. IBM Qiskit is an open-source software development kit
 that implements this circuit model of computation~\cite{Qiskit:online}. Indeed, a lot of quantum computing software for scientific applications (like quantum
 chemistry), finance, and machine learning has been developed on this platform. As such platforms become more prevalent, the number and complexity of ``quantum'' bugs increases (see Section~\ref{sec:qbp}).

We now provide a brief review of existing bugs, fixes, and bug-fix patterns in both classical and quantum programs. 

\subsection{Classical Bug and Fix Patterns}
Bug-fix patterns for classical programs are widely studied in the literature. 

Pan et al.~\cite{pan2009toward} identify 27 bug-fix patterns based on an analysis of historical bug-fix pairs. Campos and Maia~\cite{campos2017common} conduct an empirical analysis to characterize bug-fix patterns in Java open-source repositories. Soto et al.~\cite{soto2016deeper} leverage lessons learned in C projects and perform a large-scale study of bug-fix patterns of Java projects on GitHub.

Automated approaches for detecting certain bug-fix patterns are considered more efficient compared to manual approaches in general. Madeiral et al.~\cite{Madeiral2018} manually analyze hundreds of bugs and fixes and propose an automated tool based on AST to detect repair patterns in bug-fix pairs using the GumTree algorithm. The AST method is an effective solution in terms of detecting code differences. Other AST-based automated bug detection tools include~\cite{islam2020bugs,martinez2013automatically}.

\subsection{Quantum Software Engineering, Bugs and Fix Patterns}\label{sec:qbp}
With the increasing complexity of the quantum software development platforms and the programs themselves, quantum software engineering
has been receiving more attention recently. 
Indeed, the literature on testing and debugging quantum programs is growing. While quantum programs are challenging to test because of the underlying principles of quantum mechanics~\cite{miranskyy2019testing}, software engineering principles are successfully being applied to quantum program testing and debugging~\cite{miranskyy2019testing,miransky2020bug}.
On one hand, tools for analyzing the quantum software stack have been developed. For 
example, QDiff is such a tool, especially to test optimizing 
compilers that translate a quantum program to quantum gates~\cite{QDiff}. 
This is done by generating semantically equivalent quantum programs, identifying appropriate subsets, and comparing their output. Given the ``indeterminacy'' of quantum computation, they
suggest techniques for estimating the required number of 
measurements for a reliable comparison.
On the other hand, tools like QuCAT explore the testing of quantum programs through the generation of increasing strengths of combinatorial test suites until a quantum bug is isolated~\cite{QuCAT}.
See~\cite{zhao2020quantum} for a comprehensive overview of quantum software engineering research work.

Multiple approaches exist to tackle the challenge. Testing quantum programs may be simplified by adding assertion checks to the code~\cite{ali2021assessing,huang2019statistical,li2020projection,liu2020quantum} or, in some cases, introducing debugging tricks, such as extracting classical information~\cite{miransky2020bug}. We can also adapt classical fuzzy testing techniques~\cite{wang2018quanfuzz} or perform property-based testing~\cite{honarvar2020property}. The identification of bug patterns in quantum programs can assist in defect analysis and categorization~\cite{luo2022comprehensive,zhao2021identifying}. Zhao et al.~\cite{zhao2021bugs4q} propose a benchmark to evaluate testing and debugging methods for Qiskit programs. As the research of quantum software engineering is still in its infancy, the literature lacks automated solutions to detect bug-fix patterns in quantum programs. 

\section{Quantum Bug-Fix Pattern Research Agenda}\label{sec:method}

\begin{figure}[!thp]
\centering
\includegraphics[width=\columnwidth]{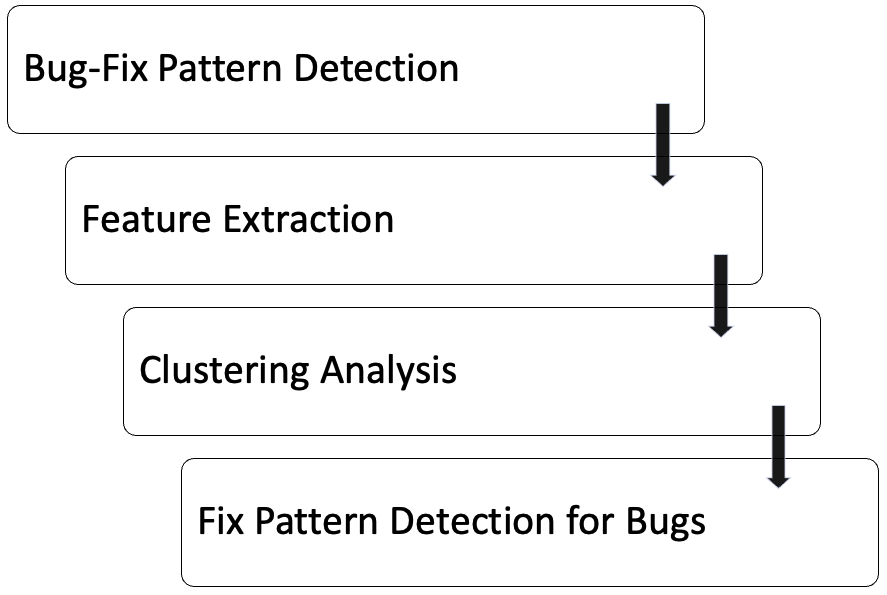}
\caption{The research agenda for detecting fix patterns for bugs. Step 1: Develop an automated tool to detect bug-fix patterns. Step 2: Extract features of bug-fix patterns. Step 3: Employ clustering algorithms for context analysis of bugs and fix patterns. Step 4: Deploy neural networks to predict fix patterns for bugs.}
\label{fig:Q-Repair}
\end{figure}

Modern software systems can be extremely complex depending on the size of the projects and the duration of the development. Quantum software will also see a similar increase in complexity as quantum hardware develops. Thus, it is critical for quantum software developers to utilize automated tools to improve their productivity and the quality of the software. Our ultimate goal is to automatically find fixes for existing bugs in quantum programs. To achieve this goal, we propose an automated method called \emph{Q-Repair} for quantum software practitioners with the following steps (illustrated in Figure~\ref{fig:Q-Repair}). 
The first step is to develop an automated tool to detect bug-fix patterns in quantum programs. The second step is to extract features of bug-fix patterns for clustering. The third step is to employ clustering algorithms for context analysis of bugs and fix patterns. The last step is the deploy artificial neural networks (ANNs) to predict fix patterns for existing quantum bugs. The detailed procedure of each step can be found below.

\begin{enumerate}
\item Develop a bug-fix pattern detection framework (e.g., \emph{Q-PAC}). At the beginning of this step, developers can manually collect bug-fix pairs from quantum development platforms (e.g., Qiskit and Q\#) and identify their patterns (e.g., incorrect initialization or measurements). Then, developers can employ syntactic and semantic analysis approaches (e.g., ASTs and regexes) to categorize bug-fix patterns automatically. The outcomes of this step (i.e., bug-fix pairs and analytical results) can be used for the next step to extract the context of bugs and fixes. 
\item Analyze features for each bug-fix pattern. From Step 1, we have collected bug-fix pairs for each category. Moreover, developers should have found and validated the syntactic and semantic features for each category using the automated framework. Thus, we focus on abstracting all the key features for each pattern, which will be used to create input datasets for the next step.
\item Use clustering algorithms (e.g., K-means) to analyze the context around the particular buggy code to find possible fixes. The goal of this step is to analyze and validate the context for each bug and fix pattern, respectively. In addition, we would like to establish the connection between clusters of buggy code and clusters of fixes. In the literature, the K-means clustering algorithm is widely used to partition observations based on the similarity of their features (or attributes). The analytical results from K-means will help us prepare the training datasets for the next step. 
\item For the final step, the goal is to predict possible fix patterns for bugs. Thus, developers can leverage ANNs to find hidden patterns and use them to identify possible remedies for bugs. One can follow the steps below to train, test, and deploy ANNs for fix pattern prediction.
\begin{enumerate}
\item Prepare a dataset for training and testing ANNs. The dataset contains buggy codes with fixes (as labels). Due to the sizes of existing open-source quantum projects, the dataset may contain both labeled and unlabeled observations (buggy code). 
\item Train ANNs with buggy code and fix patterns. For a dataset with both labeled and unlabeled data, developers can choose semi-supervised learning approaches. In semi-supervised learning, ANNs will be trained on a small set of labeled data and produce pseudo-labels. The true labeled data and pseudo-labeled data will be combined to train and improve the model. 
\item Test the performance of the models against the preliminary analysis using clustering algorithms. Developers can also use cross-validation for performance evaluation for ANNs. Cross-validation is a popular approach for training ANNs with small datasets. It partitions the dataset into subsets, trains ANNs on some of the subsets, and evaluates ANNs on the remaining ones. Cross-validation works in iterations so that training subsets and the testing subsets rotate. 
\item Deploy the best model(s) for fix pattern prediction with buggy code. Now, the ANN model is ready for fix pattern identification. Developers can feed the model with real-word and handcrafted buggy code snippets, and the model will automatically predict fix suggestions for developers.  
\end{enumerate}
\end{enumerate}

As can be seen from the steps above, the key of Step 1 is to develop an automated bug-fix pattern detection framework. In the next section, we will discuss our method to create the framework---\emph{Q-PAC}.

\section{Bug-Fix Pattern Identification with \emph{Q-PAC}}\label{sec:framework}

\begin{figure}[!thp]
\centering
\includegraphics[width=\columnwidth]{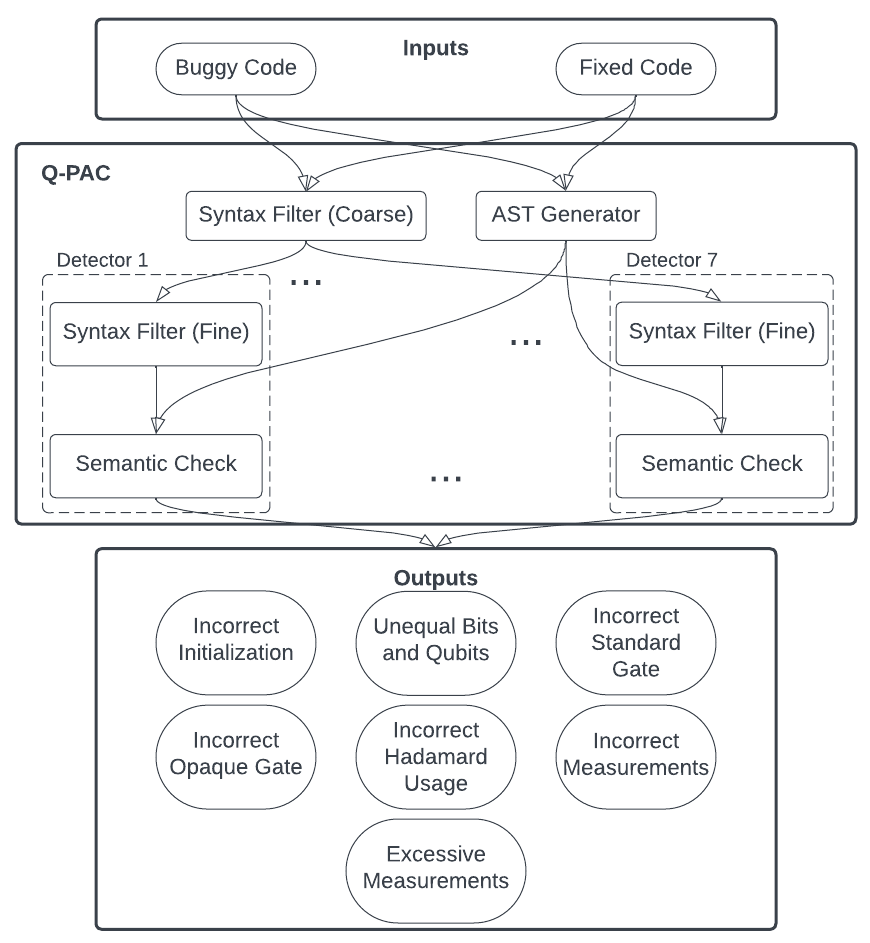}
\caption{The architecture of \emph{Q-PAC}. \textmd{Q-PAC consists primarily of four t   ypes of components: a coarse filter, an AST generator, a fine filter, and a semantic check. Code pairs are sent in parallel into the coarse filter and AST generator. The coarse filter determines which of the detectors ends up being called. Once the coarse filter selects detectors, the semantic checks (and in some cases, fine filters) corresponding to said detectors are called. The fine filters take as input the code pair. The semantic checks take ASTs as inputs from the AST generator, as well as a dictionary containing information about the program created by the fine filter (if called prior). It parses the AST, supplementing its logic with the dictionary from the fine filters where necessary, and performs a binary classification (for each detector to determine whether the current bug-fix pair falls into the pattern of the detector).}}
\label{fig:Q-PAC}
\end{figure}

Code-diffing is a technique for comparing two versions of code to identify differences or changes. This is typically done by comparing the raw text of the buggy and the fixed code, line-by-line, and identifying added, deleted, or modified lines. In this paper, we adopt two methods to collect useful information from the source code. The first method is to use ASTs~\cite{AST} to transform both the buggy and the fixed code into tree structures, and the second method is to deploy two syntax filters (in coarse and fine manners) with regex to generate dictionaries containing syntactic information.

Figure~\ref{fig:Q-PAC} shows the architecture of our tool \emph{Q-PAC}. \emph{Q-PAC} uses both methods mentioned prior, of AST extraction in combination with coarse-and-fine filters. The code files get fed into an AST generator, creating ASTs to be used later. At the same level, the \emph{coarse} filters use the buggy and fixed code files to perform a high-level classification.  Based on the coarse filter classification results, individual detectors get called. Each detector consists of either a \emph{fine} filter followed by a semantic check or just a standalone semantic check. The fine filter extracts syntactic information from the code file, and the semantic check extracts semantic information from the ASTs. Q-PAC then uses this extracted information to make a determination about that specific detector's pattern. The fine filters act to extract information that would be difficult to pull out through AST traversal alone. In cases where this information is easy to extract via a single walk through the AST, we bypass the fine filters altogether.

Currently, \emph{Q-PAC} can detect seven bug-fix patterns, that is, 1) \emph{Incorrect Initialization}, 2) \emph{Unequal Classical and Quantum Bits}, 3) \emph{Incorrect Standard Quantum Gates}, 4) \emph{Incorrect Opaque Gates}, 5) \emph{Incorrect Hadamard Gate Usage}, 6) \emph{Incorrect Measurements}, and 7) \emph{Excessive Measurements}, 

as can be seen in Figure~\ref{fig:Q-PAC}.  These seven bug patterns fall into three categories: 1) initialization-related bugs, 2) operation-related bugs, and 3) measurement-related bugs, as can be seen in Table~\ref{tab:pattern}. Note that each bug-fix pattern has its own detector, and each detector has its own logic that determines whether a bug-fix code pair falls into this category or not (details in Section~\ref{sec:qdiff}). In other words, seven detectors, corresponding to the seven bug-fix patterns, have been implemented in \emph{Q-PAC} at present. However, the sequence of logic in each detector is the same, as depicted in Figure~\ref{fig:Q-PAC}. This code structure is designed to achieve scalability because it can be extended to add more detectors and to change or augment the logic of various detectors if necessary.

There are two advantages of the current implementation of multiple, separate detectors. First, each detector can run in parallel, in principle. Second, there may exist hybrid bug-fix patterns where a bug-fix pair can be categorized into multiple patterns. In such a case, the detectors will not interfere with each other. 

Starting with a coarse categorization into the three broad classes of patterns, we descend to finer grains of categorization, ending
in detectors for specific bug-fix patterns at the ``leaf'' level. We believe that this approach has the potential to improve search efficiency substantially whenever large categories can be pruned out. We now discuss how our framework operates in detail.

\subsection{Creation of Bug-Fix Pattern List}
We identify bug-fix patterns from various sources, including the studies of~\cite{luo2022comprehensive,zhao2021identifying,zhao2021bugs4q}, StackOverflow, and GitHub repositories.\footnote{Some bug-fix patterns that we manually detect can be found at~\url{https://github.com/pranavknayak/Q-PAC/blob/main/QSEBugFindings.xlsx}} Based on our initial study, we create representative examples of commonly occurring buggy code for each of the bug-fix patterns, derived from real-world examples. We then patch them up manually and pass the code pair to \emph{Q-PAC}. This collection of representative examples is grouped into sets for each detector, with an equal number of positive examples (that match the pattern), and negative ones (that do not match the pattern), as a preliminary test suite.

\begin{table*}[!thp]
    \centering
    \caption{The seven quantum bug-fix patterns that can be detected by \emph{Q-PAC}.}
    \label{tab:pattern}
    \begin{tabular}{@{}llp{0.465 \linewidth}@{}}
    \toprule
    \emph{Bug-Fix Category} & 
    \emph{Bug-Fix Pattern} & \emph{Description} \\
    \midrule
    Initialization & Incorrect Initialization & Circuits consist of and gates act on the wrong qubits\\ 
    & Unequal Classical and Quantum Bits & There are not enough bits to store the results of measuring qubits\\
    Operation (Unitary) & Incorrect Standard Quantum Gates & The standard gate used does not perform the intended operation\\ 
     & Incorrect Opaque Gates & An opaque gate is used in place of a composite gate  \\
     & Incorrect Hadamard Gate Usage & Hadamard operations are insufficiently inverted \\
    Measurement & Incorrect Measurements & The wrong qubit(s) is/are measured  \\ 
     & Excessive Measurements &  An excessive number of measure operations causes instability  \\
    \bottomrule
    \end{tabular}
\end{table*}
\subsection{Coarse Syntax Filtering}
We group the detectors into high-level classes. This allows the detector to perform a coarse, syntax-based filtering, where we search the raw buggy files using regex that match those constructs that could be the source of individual bugs in the bug-fix patterns. In other words, the coarse filter does not determine the bug-fix pattern. Instead, it performs an initial syntactic check to determine if \emph{Q-PAC} can handle the input bug-fix pair. During the syntactic check, the absence of a sequence matching the regex means that \emph{Q-PAC} does not attempt to dissect the buggy and fixed code files further. On the other hand, the presence of a sequence matching means that further analysis is necessary, and individual detectors corresponding to that class are called. For example, we perform a sequence matching using regex\footnote{We use the Python syntax for regular expressions.} \fbox{.+measure.*} to detect if a pair needs to be passed to the detector for patterns related to measurements.

\subsection{AST Extraction} 

\begin{lstlisting}[language=Python, label={lst:ast_extract}, caption={Example of AST extraction via the \code{ast} module in Python}]
# Code
    code = "qc = QuantumCircuit(3, 3)"
# AST Generation
    print(ast.dump(ast.parse(code), indent=2))
# Output
    Module(
      body=[
        Assign(
          targets=[
            Name(id='qc', ctx=Store())],
          value=Call(
            func=Name(id='QuantumCircuit', ctx=Load()),
            args=[
              Constant(value=3),
              Constant(value=3)],
            keywords=[]))],
      type_ignores=[])
\end{lstlisting}

Besides coarse filtering, we also use ASTs to abstract code structure. Using ASTs, we first extract semantic information from the buggy and fixed code, respectively. The ASTs help abstract program semantics, allowing for more detailed analysis that is not subject to a specific organization of text within the code file. Information that ASTs provide includes identifiers as well as data about those attributes that correspond to quantum circuit objects. This information gets passed to the semantic checks for further analysis.

For example, Listing~\ref{lst:ast_extract} demonstrates an instance of AST generation, as well as what the resulting AST looks like when printed. Note the deep nesting of certain nodes within the tree, such as the identifier \code{qc} in line 10. It is this deep nesting of certain nodes that motivates the existence of fine filters.

\subsection{Bug-Fix Pattern Detectors}
The implementation of the detectors varies from one bug-fix pattern to another. As discussed at the beginning of Section~\ref{sec:framework}, we present seven detectors for seven different bug-fix patterns in this paper (details of the implementation will be discussed in Section~\ref{sec:qdiff}). All but two detectors follow a two-step process of fine filters followed by semantic checks to identify bug-fix patterns. The detectors for the \emph{Incorrect Opaque Gates} and \emph{Excessive Measurement} patterns rely only on semantic checks (see Section~\ref{sec:qdiff} for further details).

\subsubsection{Fine Syntax Filtering} 
For those bug-fix patterns that require syntactic filters, we formulate a regex. The regex is used to identify the lines of code that are relevant to a particular bug-fix pattern, and extract relevant information from those lines. If a match is found on a line of code, then we move on to the semantic check phase. When all lines of code are exhausted without finding a match, we declare that the buggy-fix code pair under investigation does not belong to the current pattern.

\subsubsection{Semantic Checks} 
After regex matching (or in its absence), we perform semantic checks. 
These checks would be specific to the bug-fix pattern detector.
For example, in the context of incorrect gates, we are still not sure if a line of code matched by the regex contains an object of quantum circuits. Thus, we analyze the information extracted from the ASTs to decide this.

\section{Implementation of \emph{Q-PAC}}\label{sec:qdiff}
In this section, we discuss examples of bug-fix patterns and their detection. Note that the examples described here are selected to illustrate the working of \emph{Q-PAC}. We provide more examples on our GitHub repository. 

As discussed in Section~\ref{sec:framework}, the bug-fix patterns that we describe here are 1) \emph{Incorrect Initialization}, 2) \emph{Unequal Classical and Quantum Bits}, 3) \emph{Incorrect Standard Quantum Gates}, 4) \emph{Incorrect Opaque Gates}, 5) \emph{Incorrect Hadamard Gate Usage}, 6) \emph{Incorrect Measurements}, and 7) \emph{Excessive Measurements}. These seven patterns correspond to the three key elements in quantum computation, i.e., 1) initialization, 2) operations, and 3) measurements. We emphasize that \emph{Q-PAC} is capable of classifying bugs that are not localized to a single line. For example, identifying an unequal number of bits and qubits is a bug that can originate multiple lines before the point where the bug gets reported. We now look at detectors for the seven bug-fix patterns with examples.

\subsection{Incorrect Initialization}\label{sec:init}

\emph{Q-PAC} detects two cases of incorrect initialization: 
1) the qubit that a gate operates on differs between the buggy and fixed codes, and 2) the number of qubits that a quantum circuit operates on differs between the buggy and fixed codes.

\subsubsection{Fine Syntax Filtering}
We use regex \fbox{.+\textbackslash..*} and \fbox{.+QuantumCircuit.*} to retrieve the lines of the code where a Qiskit gate has been used and store the names of the gates as well as the qubit indices that the gate operates on.

\subsubsection{Semantic Checks}
The detector conducts a two-step semantic check.  First, it identifies the lines where the same kind of gate is being used, to separate it from cases of the \emph{Incorrect Standard Quantum Gates} pattern. Next, it checks each of these valid gates to see if there is any difference in the qubit indices in the bug-fix code pair. If there is any difference, the detector then reports the presence of the \emph{Incorrect Initialization} pattern.

\subsubsection{Examples}

\begin{lstlisting}[language=Python, label={lst:init}, caption={Incorrect initialization}]
Buggy Code:
    qc = QuantumCircuit(2)
-   qc.h(0)
    ...
Fixed Code:
    qc = QuantumCircuit(2)
+   qc.h(1)
    ...
\end{lstlisting}

Listing~\ref{lst:init} simulates a scenario where an incorrect qubit is initialized---the developer wants to initialize the second qubit instead of the first one, which is semantically different, although syntactically correct. The only error here is in the argument to the Hadamard gate being applied. Our semantic checks identify the bug-fix and correctly classify it as an instance of the \emph{Incorrect Initialization} pattern.

\subsection{Unequal Classical and Quantum Bits}

When creating a \code{QuantumCircuit} in Qiskit, the developer can specify the number of qubits and classical bits to be associated with the circuit. This can be done in two ways: either through directly passing integer literals to the \code{QuantumCircuit} constructor or, as is more common, using a \code{QuantumRegister} and a \code{ClassicalRegister} with a pre-specified number of qubits and bits, respectively.

Herein lies the source of a very common bug, one where the number of bits and qubits within a quantum circuit do not match. If the number of bits exceeds the number of qubits, it is an inefficient allocation of resources, and reporting results tends to be messy. 

If, on the other hand, the number of qubits exceeds the number of bits, bugs arise during measurements. For program integrity, it is good practice to measure every qubit, and attempting to measure every qubit without sufficient classical bits to store the measures can result in a \code{CircuitError: Index out of range}. 

The patch for this bug is simply to ensure that quantum circuits get instantiated with an equal number of bits and qubits. 

\subsubsection{Fine Syntax Filtering}

We use two regexes, i.e.,  \fbox{.+ClassicalRegister.*} and \fbox{.+QuantumRegister.*}, to extract every such register and cache the number of bits or qubits associated with each, across buggy and patched code files.

\subsubsection{Semantic Checks}

First, we search through the ASTs for those nodes that correspond to \code{QuantumCircuit} objects, caching their names in memory for both buggy and fixed files. While caching them, we also take note of the arguments passed to their constructors and update certain counter variables. These counter variables keep track of the number of qubits and bits associated with each quantum circuit object, and they get updated by referring to the data for the registers that were extracted by the fine syntax filter. In case the \code{QuantumCircuit} objects were instantiated using integer literals, the counters just store them directly instead of referring to the register data.

Once every circuit object has been accounted for, the qubit and bit counts are compared for each one, and if there is an object that has an unequal number in the buggy code and has an equal number in the patched code, the pattern is reported by the detector. 

Simply changing the number of bits or qubits across registers will not cause this pattern to be detected by \emph{Q-PAC}, since the logic accounts specifically for those registers used when creating circuits.

\subsubsection{Examples}

In Listing~\ref{lst:unequalcbqb}, attempting to measure all three qubits results in a bug, since the developer has assumed that there are sufficient classical bits to store the measurement results in. The simple fix for this involves increasing the number of classical bits within the classical register used when creating the circuit such that it matches the number of qubits associated with the circuit.

\begin{lstlisting}[language=Python, label={lst:unequalcbqb}, caption={Example of Unequal Classical Bits and Qubits}]
Buggy Code:
    qreg = QuantumRegister(3)
-   creg = ClassicalRegister(2)
    qc = QuantumCircuit(qreg, creg)

    qc.h(0)
    qc.cx(0, 1)
    qc.cx(1, 2)
    qc.measure([0, 1, 2]. [0, 1, 2])
    ...
Fixed Code:
    qreg = QuantumRegister(3)
+   creg = ClassicalRegister(3)
    qc = QuantumCircuit(qreg, creg)
    qc.h(0)
    qc.cx(0, 1)
    qc.cx(1, 2)
    qc.measure([0, 1, 2]. [0, 1, 2])
    ...
\end{lstlisting}

\subsection{Incorrect Standard Quantum Gates}\label{sec:gate}
We now show instances where an incorrect gate is applied on a qubit of a \code{QuantumCircuit}. 

\subsubsection{Fine Syntax Filtering}\label{sec:gate-regex}
The regex for this type of pattern is \fbox{.+\textbackslash..*}, which is the same as the first one that we use for \emph{Incorrect Initialization}. However, the semantic check will be different, as will be explained in in Section~\ref{sec:gate-semantic}. The regex abstracts lines of code involving a quantum gate.  After identifying the gate operations, we will identify the names of the quantum circuits and gates for comparison.

\subsubsection{Semantic Checks}\label{sec:gate-semantic}
After the syntax checks from regex, we perform additional semantic checks in two steps. First, we check 1) if the identifiers are not equal, and 2) if they both actually belong to the inbuilt gates available in Qiskit. If either of these checks fails, we declare the bug-fix pair does not belong to \emph{Incorrect Standard Quantum Gate}. If both conditions are satisfied, \emph{Q-PAC} classifies the bug-fix pair in this category (as shown in Listing~\ref{lst:gate1}). Note that Q-PAC identifies \code{QuantumCircuit} objects independent of their actual name in the code. Listing~\ref{lst:gate2} illustrates this case.

\subsubsection{Examples}\label{sec:gate-example}

\begin{lstlisting}[language=Python, label={lst:gate1}, caption={First example of incorrect gate}]
Buggy Code:
    qc = QuantumCircuit(2)
    circuit.h(0)
-   qc.h(1)
    ...
Fixed Code:
    qc = QuantumCircuit(2)
    circuit.h(0)
+   qc.x(1)
    ...
\end{lstlisting}

\begin{lstlisting}[language=Python, label={lst:gate2}, caption={Second example of incorrect gate}]
Buggy Code:
-   a = QuantumCircuit(2)
-   a.sdg(1)
    ...
Fixed Code:
+   qc = QuantumCircuit(2)
+   qc.tdg(1)
    ...
\end{lstlisting}

Listing~\ref{lst:gate1} (derived from Stack Overflow~\cite{python2e58:online}) illustrates a scenario where the gate operation is incorrect in \texttt{\code{QuantumCircuit}}. The identifiers are \code{h} and \code{x} in this example, both of which correspond to valid inbuilt gates in Qiskit, namely, the \texttt{Hadamard} gate and the \texttt{X}-gate. Since the type of gate applied to the circuit is changed, \emph{Q-PAC} classifies this code pair as belonging to the \emph{Incorrect Standard Quantum Gates} pattern.

Listing~\ref{lst:gate2} is very similar to the first example, except that now the \code{QuantumCircuit} object has different names, that is, \code{a} and \code{qc}. 
In the second example, we first check if the underlying gate is being accessed by a \code{QuantumCircuit} object. We do this using the data extracted from the AST of the buggy and the fixed code. This removes the dependency on the identifier name to determine if a bug-fix pair belongs to this pattern or not.

\subsection{Incorrect Opaque Gates}\label{sec:opaque}

We choose to only perform semantic checks for this pattern and skip the syntactic filter stage because all the information required to make a determination can be extracted from a single pass through the AST.

When working with advanced circuits in Qiskit, there often comes the need to manipulate more than two qubits. This can cause \emph{Incorrect Opaque Gates} pattern to arise, in which the developer tries to create an opaque gate that attempts to control more than two qubits. To fix this, a composite gate is needed--one that acts on more than two quantum wires but is
constructed using basic gates (both built-in and opaque gates) that only act on two or fewer qubits. 

\subsubsection{Semantic Checks}
We detect occurrences of this pattern as follows. 

First, we check for all opaque gate instantiations that attempt to control three or more qubits across both buggy and patched abstract syntax trees. We cache the names of each of these opaque gates in memory.

Second, we search through the ASTs to find all method calls that wrap a \code{QuantumCircuit} object into a composite gate by caching the targets of any assignment that results from a \code{to\_instruction()} method call.  

Lastly, we compare the total counts of opaque gates and \code{to\_instruction()} calls across buggy and patched files. We check if any decrease in the count of opaque gate instances is matched or exceeded by a corresponding increase in the count of composite gate instances. We check across counts instead of comparing on a gate-by-gate basis. This is because we believe that the fix for the bug spans enough lines that the names of the gates need not persist. This means that no assumptions are made about the names of the gates across buggy and fixed files, and the detector only takes the number of gates into consideration. While a matching change in opaque and composite gates should report the pattern as detected, we also allow for an increase in composite gates that exceeds the decrease in violating opaque gates to count as an instance of the pattern. This is because we leave room for composite gates to be created outside of the patch.

\subsubsection{Examples}
\begin{lstlisting}[language=Python, label={lst:opaque}, caption={Example of opaque gate to composite gate conversion}]
Buggy Code:
    qc = QuantumCircuit(3, 3)
-   gt = Gate('my_custom_gate', 3, [])
    qc.append(gt, [0, 1, 2])
    ...
Fixed Code:
    qc = QuantumCircuit(3, 3)
+   sub_circuit = QuantumCircuit(3, name='sub_circ')
+   ... # Apply basic gates to sub_circuit to emulate gt's behaviour
+   gt = sub_circuit.to_instruction()
    qc.append(gt, [0, 1, 2])
    ...
\end{lstlisting}

In Listing~\ref{lst:opaque}, the source of the bug comes from instantiating an opaque gate object that attempts to directly control three qubits. The fix for this involves creating a quantum sub-circuit with three quantum wires (one for each of the qubits we need to control) and applying one- or two-qubit gates in composition such that the net effect is the same as what was desired from the original opaque gate. Once that sub-circuit has been created, it can be wrapped into a composite gate object by calling the circuit's \code{to\_instruction()} method, allowing it to be appended to a larger circuit, emulating the properties of a gate.

\subsection{Incorrect Hadamard Gate Usage}
Some quantum algorithms require the basis to be rotated before some gates are applied and then rotated back. Examples include the standard approach to the Hidden Subgroup Problem~\cite{lomont2004hidden}, where the basis is rotated from the computational to Fourier and back. As a simple example, consider the Deutsch-Jozsa algorithm, where multiple Hadamard (Fourier) gates are applied along a quantum wire~\cite{DJ}. More specifically, the Hadamard gate applied must also be inverted further along the wire (by applying the Hadamard gate once again, because it is its own inverse). Within this context, a bug could arise in case every such Hadamard is not matched with an inverse Hadamard transformation along a wire. A straightforward fix for this is to simply remove or add Hadamard gates along each qubit's wire until every qubit has an even number of these gates
applied on it.

For this specific pattern's implementation, the order of data flow reverses, with the semantic checks extracting information and creating data structures that are then fed into the fine syntactic filters. However, both syntactic and semantic checks remain within the same layer of the call stack.

\subsubsection{Semantic Checks}
The semantic checks follow a similar logic to the previous six detectors. We traverse the buggy and patched ASTs and cache all \code{QuantumCircuit} objects. These cached objects are stored in data structures that keep track of the number of Hadamard gates applied to each unique qubit within a circuit. The semantic checks then feed these data structures to the syntactic check so that the actual Hadamard counts can be populated.
\subsubsection{Fine Syntax Filtering}
The syntactic checks use two different regexes, each of which has a specific purpose. The first is the regex \fbox{\textbackslash .h(.* )}, which we call the Qubit regex. The syntactic check uses it for two purposes. The first is to access the specific qubit of the circuit that gets indexed by Qiskit's Hadamard method call, and the second is as an indicator that a Hadamard gate has been encountered and that the count for that qubit must get updated. The second regex we use is \fbox{.+\textbackslash.h}, the Circuit regex, which acts as a filter that extracts the name of the circuit that the qubit under consideration exists in. We use these two to index into the data structure passed down by the semantic check and update the number of Hadamard gates encountered by 1. 

Finally, we check to see if, for any qubit, the number of Hadamards performed changed from even to odd, in which case, \emph{Q-PAC} reports that the pattern has been encountered.
\subsubsection{Examples}

\begin{lstlisting}[language=Python, label={lst:hadamard}, caption={Example of Incorrect Hadamard Gate Usage}]
Buggy Code:
    qc = QuantumCircuit(3, 3)
>   qc.h(0)
>>  qc.h(1)
    ...
>>  qc.h(1)
    ...
Fixed Code:
    qc = QuantumCircuit(3, 3)
>   qc.h(0)
>>  qc.h(1)
    ...
>>  qc.h(1)
    ...
>   qc.h(0)
\end{lstlisting}
In Listing~\ref{lst:hadamard}, we present a simplified example where all the details present between any two Hadamard applications are abstracted away. In this example, the buggy code has only a single gate acting on the first qubit of the circuit (line number 3), while in the patched code, that qubit also has the gate's inverse applied at the end (line number 15). The highlighted lines show the corresponding pairs of Hadamards and their inversions.

\subsection{Incorrect Measurements}\label{sec:measure}

A very common source of bugs in quantum software is the incorrect application of measurements. A measurement could be incorrect in a variety of ways, from measuring the wrong qubit, to measuring the right qubit but at the wrong point in a circuit, to storing the results of measurements in the wrong classical bits. A broad fix for all of these is to change the order of application of measurements relative to other operations in the circuit, and to change the arguments passed to the \code{measure} call. Simply changing the number of measurements can also be a potential fix for this class of bugs.

\subsubsection{Fine Syntax Filtering}
The fine syntactic filter in this case uses the regex \fbox{.+\textbackslash.measure.*} It passes down to the semantic checks a dictionary containing every measure operation in the code as well as the lines they occur in. This will be used to determine changes in the ordering of measures within a circuit.

\subsubsection{Semantic Checks}

The semantic checks perform analysis for three commonly encountered \code{measure} functions exist in Qiskit, namely, \code{measure()}, \code{measure\_all()}, and \code{measure\_inactive()}. It first computes the total counts of all variants of \code{measure} functions used in the buggy and fixed code, respectively. If the totals are different, the bug-fix pair is classified within this pattern; otherwise, \emph{Q-PAC} will go to the next step. Note that we assume multiple \code{measure} functions have a one-to-one correspondence across buggy and fixed code.

\emph{Q-PAC} then checks if the counts and order of individual measure functions stays the same across buggy and patched codes. If not, then the detector reports the presence of the \emph{Incorrect Measurements} pattern; otherwise, \emph{Q-PAC} performs further analysis.

In the next stage of analysis, the detector checks if the arguments passed to the \code{measure} functions are the same at the same lines in the buggy and patched codes.. Since arguments are usually qubit lists, we use \code{numpy} arrays to check the difference. If the arguments are different, then this pattern is reported; otherwise, \emph{Q-PAC} will go to the last step of the semantic check.

Finally, if both the set of \code{measure} functions and the arguments passed to them are the same, \emph{Q-PAC} comapres their positions. If corresponding measurements are performed at different lines in across buggy and patched codes, this code pair is classified as belonging to the this pattern.

\subsubsection{Examples}

\begin{lstlisting}[language=Python, label={lst:measure1}, caption={First example of incorrect measurement}]
Buggy Code:
    qr = QuantumRegister(2, name='qreg')
    cr = ClassicalRegister(2, name='creg')
    qc = QuantumCircuit(qr,cr)
    qc.h(qr)
-   qc.measure_all()
Fixed Code:
    qr = QuantumRegister(2, name='qreg')
    cr = ClassicalRegister(2, name='creg')
    qc = QuantumCircuit(qr,cr)
    qc.h(qr)
+   qc.measure(qc.qubits, qc.clbits)
\end{lstlisting}

We show two examples here. In Listing~\ref{lst:measure1} (derived from \code{qiskit-terra} issue report \#6751), though valid \code{measure} functions are used in both bug and fix, \code{measure\_all()} function shows in ``bug'' while \code{measure()} function is its replacement in ``fix''. The outputs of the \code{measure\_all()} function and the \code{measure()} function are not expected to be the same, indicating the presence of the \emph{Incorrect Measurements} pattern.
\begin{lstlisting}[language=Python, label={lst:measure2}, caption={Second example of incorrect measurement}]
Buggy Code:
    qc = QuantumCircuit(3,3)
    qc.x(0)
    qc.barrier()
-   qc.measure([0,1,2],[0,1,2])
Fixed Code:
    qc = QuantumCircuit(3,3)
    qc.x(0)
    qc.barrier()
+   qc.measure([0,1,2],[1,0,2]) 
\end{lstlisting}

Similar to Listing~\ref{lst:measure1}, although valid measure functions are being used in Listing~\ref{lst:measure2} (derived from \code{qiskit-aer} issue report \#664), the outputs written to the classical bits are in the wrong order in the buggy code. Hence, this is an instance of \emph{Incorrect Measurement}, and \emph{Q-PAC} classifies it as such.

\subsection{Excessive Measurements}
While there is no syntactic limit or logical constraint on performing measurements on a quantum circuit, this bug tends to arise from instability in simulator backends. 

For example, when using the Qiskit Aer simulator backend, performing repeated measurements on a circuit can result in conflicting reports on the number of qubits, as well as certain methods simply not running as expected~\cite{zhao2021identifying}.  

These issues mean that it is recommended practice to minimize excessive operations on qubits. In case these bugs tend to arise due to excessive measurement, the straightforward patch is to simply reduce the number of such operations.

\subsubsection{Semantic Checks}
As with the opaque gate pattern, we skip the syntactic checks here because all the information required to make a determination can be extracted from a single walk through the abstract syntax trees. 

We traverse the ASTs of both buggy and patched code files, caching all instances of \code{QuantumCircuit} objects in order to track the number of measurements performed on each. Within the same pass, we make a note of any occurrence of measurements on our cached \code{QuantumCircuit}. Any measure encountered results in the counts for the corresponding circuits being updated. Note that we assume that circuits are instantiated before any measures are performed on them.

We especially account for measurements within the scope of a \code{for} loop, incrementing the counts for the circuit by the number of iterations the loop runs for. We make the assumption that any measure within the body of a loop occurs once for every iteration of the loop. \emph{Q-PAC} is capable of analyzing loops that are created by iterating through \code{range} generators or list literals. 

Once this has been performed for both code files, we compare, for each \code{QuantumCircuit} that appears across both code files, the number of measurements performed. If, for any one, the number of measurements is reduced, the detector reports that the pattern has been detected.

Note that we assume that the names of\code{QuantumCircuit} objects persists through the fix.

\subsubsection{Examples}

\begin{lstlisting}[language=Python, label={lst:excessive}, caption={Example of Excessive Measurements}]
Buggy Code:
    qreg = QuantumRegister(10)
    creg = ClassicalRegister(10)
    circ = QuantumCircuit(qreg, creg)
-   for i in range(10):
        circ.measure(qreg[i], mreg[i]
    ...
Fixed Code:
    qreg = QuantumRegister(10)
    creg = ClassicalRegister(10)
    circ = QuantumCircuit(qreg, creg)
+   for i in range(5):
        circ.measure(qreg[i], mreg[i]
    ...
\end{lstlisting}

\subsection{A ``Counter Example'' Epilogue}
How would \emph{Q-PAC} behave if there is a syntactic (or textual) difference between
the buggy and the fixed code, but the difference has no semantic consequence?

\begin{lstlisting}[language=Python, label={lst:counter1}, caption={Code identified as a correct implementation}]
Buggy Code:
    qc = QuantumCircuit(2)
-   qc.h(0+1)
    ...
Fixed Code:
    qc = QuantumCircuit(2)
+   qc.h(1)
    ...
\end{lstlisting}

In Listing~\ref{lst:counter1}, we notice that the underlying logic in both the code is exactly the same, but for the representation, wherein the buggy code, qubit $0$ is denoted by qubit $0+1$. A manual check determines this is not a bug-fix pattern, and our tool does the same---it is capable of recognizing the semantics as opposed to just reporting text differences.

However, we note that certain syntactic changes do result in misclassification by \emph{Q-PAC}. 

\begin{lstlisting}[language=Python, label={lst:counter2}, caption={Code identified as a correct implementation}]
Buggy Code:
    qc = QuantumCircuit(2)
-   qc.h(0)
-   qc.x(1)
    ...
Fixed Code:
    qc = QuantumCircuit(2)
+   qc.x(1)
+   qc.h(0)    
    ...
\end{lstlisting}

Listing~\ref{lst:counter2} is one example of such a misclassification. There is no real difference in the semantics between buggy and patched code files, since while the order of application of gates to the circuit differs within the code, each gate is being applied to independent qubits. The same circuit results from both files, but \emph{Q-PAC} detects an \emph{Incorrect Standard Quantum Gates} pattern, as a result of analyzing only the order of gates within the circuit definition for that detector. We emphasize
that such misclassifications can be avoided through more in-depth
analysis by the detectors. However, it is difficult to 
resolve inherent ambiguities--a simple example being whether to
classify two bug-fixes on different lines as independent, or
part of a larger composite bug-fix pattern. An alternative would
be to output a list of potential pattern classes and let the user choose the
appropriate one. Indeed, the current implementation of \emph{Q-PAC} follows this approach.
\section{Threats to Validity}\label{sec:threats}
Validity threats are classified according to~\cite{wohlin2012experimentation,yin2009case}. 

\textbf{Internal and construct validity:} In this initial study, we test \emph{Q-PAC} with bug-fix patterns where both one line of code and multiple lines of code are modified. We develop both positive and negative test cases to verify our implementation as proof-of-concept and will test our framework with real bug-fix code. However, there are bugs that exist in multiple files, and these bugs require fixes in multiple files. Those cases are not covered in this paper.

\textbf{External and conclusion validity:} Software engineering studies suffer from the generalization problem, which can only be solved partially~\cite{wieringa2015six}. Although we cover seven of the most significant bug-fix patterns of Qiskit in quantum computations~\cite{zhao2021identifying}, our findings may not generalize to other projects, especially quantum projects in other platforms, e.g., Q\#. Thus, we publish our source code on GitHub. Moreover, all the detectors in \emph{Q-PAC} are independent. Thus, \emph{Q-PAC} can be expanded to include more detectors and used to determine more bug-fix patterns.
Together with other bug detection techniques, such as~\cite{zhao2023qchecker,zhao2021identifying}, we also hope the quantum software engineering community will work closely and proactively to invent more novel methods and tools to improve the quality of quantum software.

\section{Conclusions and Future Work}\label{sec:conclusion}

In this paper, we propose a research agenda called \emph{Q-Repair} to automatically detect fix patterns for quantum software bugs. As an initial step, we conduct an analysis of bug-fix patterns in quantum programs. We develop a framework called \emph{Q-PAC} to automatically detect Qiskit bug-fix patterns and demonstrate its effectiveness with seven different patterns and various examples. An obvious future direction is to enhance the tool with more detectors to cover a "nearly exhaustive" list of bug-fix patterns, such as unhandled exceptions and incorrect implementations of standard quantum algorithms. It is also in our interest to proceed with the validation of the proposed \emph{Q-Repair} method, particularly with the immediate next step of employing clustering algorithms to analyze the context of buggy code and using artificial neural networks to predict fix patterns.

\bibliographystyle{abbrv}

\bibliography{references}

\begin{thebibliography}{10}

\bibitem{ali2021assessing}
S.~Ali, P.~Arcaini, X.~Wang, and T.~Yue.
\newblock Assessing the effectiveness of input and output coverage criteria for
  testing quantum programs.
\newblock In {\em 2021 14th IEEE Conference on Software Testing, Verification
  and Validation (ICST)}, pages 13--23. IEEE, 2021.

\bibitem{StatAn}
N.~Ayewah, W.~Pugh, D.~Hovemeyer, J.~D. Morgenthaler, and J.~Penix.
\newblock Using static analysis to find bugs.
\newblock {\em IEEE Software}, 25(5):22--29, 2008.

\bibitem{IncorrProg}
A.~Barr.
\newblock {\em Find the Bug: A Book of Incorrect Programs}.
\newblock Addison-Wesley Professional, 2004.

\bibitem{campos2017common}
E.~C. Campos and M.~de~Almeida~Maia.
\newblock Common bug-fix patterns: A large-scale observational study.
\newblock In {\em 2017 ACM/IEEE International Symposium on Empirical Software
  Engineering and Measurement (ESEM)}, pages 404--413. IEEE, 2017.

\bibitem{daley2022practical}
A.~J. Daley, I.~Bloch, C.~Kokail, S.~Flannigan, N.~Pearson, M.~Troyer, and
  P.~Zoller.
\newblock Practical quantum advantage in quantum simulation.
\newblock {\em Nature}, 607(7920):667--676, 2022.

\bibitem{DJ}
D.~Deutsch and R.~Jozsa.
\newblock Rapid solution of problems by quantum computation.
\newblock {\em Proc. R. Soc. Lond. A}, 439:553–558, 1992.

\bibitem{honarvar2020property}
S.~Honarvar, M.~R. Mousavi, and R.~Nagarajan.
\newblock Property-based testing of quantum programs in q\#.
\newblock In {\em Proc. of the IEEE/ACM 42nd International Conference on
  Software Engineering Workshops}, pages 430--435, 2020.

\bibitem{huang2022quantum}
H.-Y. Huang, M.~Broughton, J.~Cotler, S.~Chen, J.~Li, M.~Mohseni, H.~Neven,
  R.~Babbush, R.~Kueng, J.~Preskill, et~al.
\newblock Quantum advantage in learning from experiments.
\newblock {\em Science}, 376(6598):1182--1186, 2022.

\bibitem{huang2019statistical}
Y.~Huang and M.~Martonosi.
\newblock Statistical assertions for validating patterns and finding bugs in
  quantum programs.
\newblock In {\em Proc. of the 46th International Symposium on Computer
  Architecture}, ISCA'19, page 541–553. Association for Computing Machinery,
  2019.

\bibitem{islam2020bugs}
M.~R. Islam and M.~F. Zibran.
\newblock How bugs are fixed: Exposing bug-fix patterns with edits and nesting
  levels.
\newblock In {\em Proc. of the 35th annual ACM symposium on applied computing},
  pages 1523--1531, 2020.

\bibitem{AST}
J.~Jones.
\newblock Abstract syntax tree implementation idioms.
\newblock {\em Pattern Languages of Program Design}, 2003.
\newblock Proceedings of the 10th Conference on Pattern Languages of Programs
  (PLoP2003) http://hillside.net/plop/plop2003/papers.html.

\bibitem{Kher23}
K.~V. Kher, M.~B. Chandra, I.~Joshi, L.~Zhang, and M.~V.~P. Rao.
\newblock Automatic diagnosis of quantum software bug-fix motifs.
\newblock In {\em Proc. of the 35th International Conference on Software
  Engineering and Knowledge Engineering (SEKE23)}, 2023.

\bibitem{BugTypes}
Y.~Lee and J.~Yang.
\newblock Analysis of bug types of textbook code with open-source software.
\newblock In H.~R. Arabnia, L.~Deligiannidis, F.~G. Tinetti, and Q.-N. Tran,
  editors, {\em Advances in Software Engineering, Education, and e-Learning},
  pages 629--639, Cham, 2021. Springer International Publishing.

\bibitem{li2020projection}
G.~Li, L.~Zhou, N.~Yu, Y.~Ding, M.~Ying, and Y.~Xie.
\newblock Projection-based runtime assertions for testing and debugging quantum
  programs.
\newblock {\em Proc. of the ACM on Programming Languages},
  4({OOPSLA}):150:1--150:29, 2020.

\bibitem{liu2020quantum}
J.~Liu, G.~T. Byrd, and H.~Zhou.
\newblock Quantum circuits for dynamic runtime assertions in quantum
  computation.
\newblock In {\em Proc. of the 25th International Conference on Architectural
  Support for Programming Languages and Operating Systems}, ASPLOS'20, page
  1017–1030. Association for Computing Machinery, 2020.

\bibitem{lomont2004hidden}
C.~Lomont.
\newblock The hidden subgroup problem-review and open problems.
\newblock {\em arXiv preprint quant-ph/0411037}, 2004.

\bibitem{luo2022comprehensive}
J.~Luo, P.~Zhao, Z.~Miao, S.~Lan, and J.~Zhao.
\newblock A comprehensive study of bug fixes in quantum programs.
\newblock In {\em 2022 IEEE International Conference on Software Analysis,
  Evolution and Reengineering (SANER)}, pages 1239--1246. IEEE, 2022.

\bibitem{Madeiral2018}
F.~Madeiral, T.~Durieux, V.~Sobreira, and M.~Maia.
\newblock Towards an automated approach for bug fix pattern detection.
\newblock In {\em Proc. of the VI Workshop on Software Visualization, Evolution
  and Maintenance (VEM)}, 2018.

\bibitem{martinez2013automatically}
M.~Martinez, L.~Duchien, and M.~Monperrus.
\newblock Automatically extracting instances of code change patterns with ast
  analysis.
\newblock In {\em 2013 IEEE international conference on software maintenance},
  pages 388--391. IEEE, 2013.

\bibitem{miranskyy2019testing}
A.~Miranskyy and L.~Zhang.
\newblock On testing quantum programs.
\newblock In {\em Proc. of the 2019 IEEE/ACM 41st International Conference on
  Software Engineering: New Ideas and Emerging Results (ICSE-NIER)}, pages
  57--60. IEEE, 2019.

\bibitem{miransky2020bug}
A.~Miranskyy, L.~Zhang, and J.~Doliskani.
\newblock Is your quantum program bug-free?
\newblock In {\em Proc. of the ACM/IEEE 42nd International Conference on
  Software Engineering: New Ideas and Emerging Results}, ICSE-NIER '20, page
  29–32. ACM, 2020.

\bibitem{miranskyy2021testing}
A.~Miranskyy, L.~Zhang, and J.~Doliskani.
\newblock On testing and debugging quantum software.
\newblock {\em arXiv preprint arXiv:2103.09172}, 2021.

\bibitem{nielsen_chuang_2010}
M.~A. Nielsen and I.~L. Chuang.
\newblock {\em Quantum Computation and Quantum Information: 10th Anniversary
  Edition}.
\newblock Cambridge Univ. Press, 2010.

\bibitem{pan2009toward}
K.~Pan, S.~Kim, and E.~J. Whitehead.
\newblock Toward an understanding of bug fix patterns.
\newblock {\em Empirical Software Engineering}, 14:286--315, 2009.

\bibitem{Piattini}
M.~Piattini et~al.
\newblock The talavera manifesto for quantum software engineering and
  programming.
\newblock In {\em Proc. of the 1st International Workshop on the QuANtum
  SoftWare Engineering {\&} pRogramming, Talavera de la Reina, Spain, 2020},
  volume 2561 of {\em {CEUR} Workshop Proceedings}, pages 1--5. CEUR-WS.org,
  2020.

\bibitem{BugDataSet}
V.~Sobreira, T.~Durieux, F.~Madeiral, M.~Monperrus, and M.~Maia.
\newblock Dissection of a bug dataset: Anatomy of 395 patches from defects4j.
\newblock In {\em SANER 2018}, 03 2018.

\bibitem{soto2016deeper}
M.~Soto, F.~Thung, C.-P. Wong, C.~Le~Goues, and D.~Lo.
\newblock A deeper look into bug fixes: patterns, replacements, deletions, and
  additions.
\newblock In {\em Proc. of the 13th International Conference on Mining Software
  Repositories}, pages 512--515, 2016.

\bibitem{python2e58:online}
{Stack Overflow}.
\newblock 2 entangled qubit gives all states with 25 \%.
\newblock
  \url{https://stackoverflow.com/questions/62661255/2-entangled-qubit-gives-all-states-with-25},
  2022.

\bibitem{Qiskit:online}
M.~Treinish, J.~Gambetta, et~al.
\newblock Qiskit/qiskit: Qiskit 0.39.5.
\newblock \url{https://doi.org/10.5281/zenodo.7545230}, Jan. 2023.

\bibitem{wang2018quanfuzz}
J.~Wang, M.~Gao, Y.~Jiang, J.~Lou, Y.~Gao, D.~Zhang, and J.~Sun.
\newblock Quanfuzz: Fuzz testing of quantum program.
\newblock {\em arXiv preprint arXiv:1810.10310}, 2018.

\bibitem{QDiff}
J.~Wang, Q.~Zhang, G.~H. Xu, and M.~Kim.
\newblock Qdiff: Differential testing of quantum software stacks.
\newblock In {\em 2021 36th IEEE/ACM International Conference on Automated
  Software Engineering (ASE)}, pages 692--704, 2021.

\bibitem{QuCAT}
X.~Wang, P.~Arcaini, T.~Yue, and S.~Ali.
\newblock Qucat: {A} combinatorial testing tool for quantum software.
\newblock {\em CoRR}, abs/2309.00119, 2023.

\bibitem{wieringa2015six}
R.~J. Wieringa and M.~Daneva.
\newblock Six strategies for generalizing software engineering theories.
\newblock {\em Science of computer programming}, 101:136--152, 4 2015.

\bibitem{wohlin2012experimentation}
C.~Wohlin, P.~Runeson, M.~H{\"o}st, M.~Ohlsson, B.~Regnell, and A.~Wessl{\'e}n.
\newblock {\em Experimentation in Software Engineering}.
\newblock Computer Science. Springer Berlin Heidelberg, 2012.

\bibitem{yin2009case}
R.~Yin.
\newblock {\em Case Study Research: Design and Methods}.
\newblock Applied Social Research Methods. SAGE Publications, 2009.

\bibitem{testing}
M.~Young and M.~Pezze.
\newblock {\em Software Testing and Analysis: Process, Principles and
  Techniques}.
\newblock John Wiley \&amp; Sons, Inc., Hoboken, NJ, USA, 2005.

\bibitem{zhao2020quantum}
J.~Zhao.
\newblock Quantum software engineering: Landscapes and horizons.
\newblock {\em arXiv preprint arXiv:2007.07047}, 2020.

\bibitem{zhao2023qchecker}
P.~Zhao, X.~Wu, Z.~Li, and J.~Zhao.
\newblock Qchecker: Detecting bugs in quantum programs via static analysis.
\newblock {\em arXiv preprint arXiv:2304.04387}, 2023.

\bibitem{zhao2021identifying}
P.~Zhao, J.~Zhao, and L.~Ma.
\newblock Identifying bug patterns in quantum programs.
\newblock In {\em Proc. of the 2021 IEEE/ACM 2nd International Workshop on
  Quantum Software Engineering (Q-SE)}, pages 16--21. IEEE, 2021.

\bibitem{zhao2021bugs4q}
P.~Zhao, J.~Zhao, Z.~Miao, and S.~Lan.
\newblock Bugs4q: A benchmark of real bugs for quantum programs.
\newblock In {\em 2021 36th IEEE/ACM International Conference on Automated
  Software Engineering (ASE)}, pages 1373--1376. IEEE, 2021.

\end{thebibliography}

\end{document}